\documentclass[
  journal=largetwo,
  manuscript=article-type,
  year=2020,
  volume=37,
]{cup-journal}

\usepackage{amsmath}
\usepackage[nopatch]{microtype}
\usepackage{booktabs}

\newcommand{\fink}{{\sc Fink}}

\title{Real-Time Active Learning for optimised spectroscopic follow-up: Enhancing early SN Ia classification with the Fink broker}

\author{A. M{\"o}ller}
\affiliation{Centre for Astrophysics and Supercomputing, Swinburne University of Technology, John St, Hawthorn, VIC 3122, Australia}
\alsoaffiliation{ARC Centre of Excellence for Gravitational Wave Discovery (OzGrav), John St, Hawthorn, VIC 3122, Australia}
\email[A. Möller]{amoller@swin.edu.au}

\author{E. E. O. Ishida}
\affiliation{LPCA, Université Clermont Auvergne, CNRS/IN2P3, F-63000 Clermont-Ferrand, France}

\author{J. Peloton}
\affiliation{Université Paris-Saclay, CNRS/IN2P3, IJCLab, 91405 Orsay, France}

\author{O. Vidal Velázquez}
\affiliation{Centre for Astrophysics and Supercomputing, Swinburne University of Technology, John St, Hawthorn, VIC 3122, Australia}
\alsoaffiliation{ARC Centre of Excellence for Gravitational Wave Discovery (OzGrav), John St, Hawthorn, VIC 3122, Australia}

\author{J. Soon}
\affiliation{The Research School of Astronomy and Astrophysics, Australian National University, Cotter Rd, Weston Creek ACT 2611, Australia}

\author{B. Martin}
\affiliation{The Research School of Astronomy and Astrophysics, Australian National University, Cotter Rd, Weston Creek ACT 2611, Australia}

\author{M. Cluver}
\affiliation{Centre for Astrophysics and Supercomputing, Swinburne University of Technology, John St, Hawthorn, VIC 3122, Australia}

\author{M. Leoni}
\affiliation{Université Paris-Saclay, CNRS/IN2P3, IJCLab, 91405 Orsay, France}

\author{E. Taylor}
\affiliation{Centre for Astrophysics and Supercomputing, Swinburne University of Technology, John St, Hawthorn, VIC 3122, Australia}


\begin{document}

\begin{abstract}

Current and future surveys rely on machine learning classification to obtain large and complete samples of transients. Many of these algorithms are restricted by training samples that contain a limited number of spectroscopically confirmed events. Here, we present the first real-time application of Active Learning to optimise spectroscopic follow-up with the goal of improving training sets of early type Ia supernovae (SNe Ia) classifiers. 

Using a photometric classifier for early SN Ia, we apply an Active Learning strategy for follow-up optimisation using the real-time \fink\ broker processing of the ZTF public stream. We perform follow-up observations at the ANU 2.3m telescope in Australia and obtain 92 spectroscopic classified events that are incorporated in our training set.

We show that our follow-up strategy yields a training set that, with $25\%$ less spectra, improves classification metrics when compared to publicly reported spectra. Our strategy selects in average fainter events and, not only supernovae types, but also microlensing events and flaring stars which are usually not incorporated on training sets.

Our results confirm the effectiveness of active learning strategies to construct optimal training samples for astronomical classifiers. With the Rubin Observatory LSST soon online, we propose improvements to obtain earlier candidates and optimise follow-up. This work paves the way to the deployment of real-time AL follow-up strategies in the era of large surveys.

\end{abstract}

\section{Introduction}

Photometric classification is becoming one of the most efficient tools to harness the full power of large surveys detecting transients and variables. In the last decade thousands of supernovae have been detected by surveys such as the Dark Energy Survey and the Zwicky Transient Facility \citep[DES and ZTF;][]{Bernstein:2011,Bellm:2019}. However, only a small percentage has been spectroscopically followed-up and classified. Recently, in the Dark Energy Survey, photometric classification has allowed to identify more than 3 times more type Ia supernovae (SNe Ia) compared to spectroscopic classification for cosmology analyses \citep{DES:2024, Moller:2022}, as well as a close-to-complete sample of high-quality SNe Ia in the survey \citep{Moller:2024}.

Photometric classification (PC) uses only light-curves, or flux evolution as a function of time, together with other useful information, to obtain probabilities of a light-curve being that of a given astrophysical event. In the last decade, many PC methods have been developed to identify transients and variables. Most of them rely on Machine Leaning (ML) algorithms trained either with real data or simulations for SN classification \citep{Moller:2016,Moller:2020,Qu:2021,Boone:2021,Gagliano:2023}, kilonovae \citep{Biswas:2023}, variables \citep{Sanchez-Saez:2021} or identifying simultaneously many transients and variable classes \citep{Fraga:2024, Cabrera-Vives:2024}. However, training sets are far from complete at high redshifts and, in particular for rare and new classes, scarce. Recent efforts have focused in improving training sets through refined simulations \citep{Vincenzi:2019} or by combining real-data and simulations \citep{Carrick:2021}.

Starting in 2025, for a decade, the Legacy Survey of Space and Time (LSST) at the Vera C. Rubin Observatory will detect up to 10 million time-changing events per night, and more than a million SNe during the whole survey \citep[Rubin, ][]{LSST:2009}. It will be impossible to follow-up all transient candidates spectroscopically or even a subset of such as all type Ia supernovae (SNe Ia), SNe, nor rare transients. Thus, to harness the power of LSST, it will be necessary to develop and apply photometric classification methods to the large volumes of Rubin data.

Spearheading the effort of identifying the most promising candidates for a variety of science cases are the Rubin Community Brokers. During its ten year survey, LSST, will send in real-time all transient and variable detections as a stream to 7 community brokers: \fink\ \citep{Moller:2021}, \texttt{ALERCE} \citep{alerce}, \texttt{AMPEL} \citep{ampel}, \texttt{ANTARES} \citep{antares}, \texttt{Babamul}, \texttt{LASAIR} \citep{lasair} and \texttt{Pitt-Google}. 

The \fink\ broker, is designed to ingest and process large data streams within minutes. Currently processing the ZTF stream as a precursor, it will be deployed with Rubin LSST data from 2025. To identify the most promising candidates for a variety of science cases, it makes use of several different science modules which contain cross-matching capabilities, ML classifiers and user-specified filters \citep{Moller:2021}. An evaluation of current and future \fink\ ML algorithms performance, including an early SNIa classifier, using Rubin LSST simulations can be found in \cite{Fraga:2024}. \fink\ also is actively developing novel systems that incorporate active learning for anomaly detection and optimising follow-up resources.

Active learning (AL) has been shown to be a promising strategy for anomaly detection \citep{Ishida:2021, Lochner:2021} and to optimise training sets. Using supernova light-curve simulations from the Supernova Photometric Classification Challenge  \citep[SNPCC, ][]{Kessler:2010PC}, \cite{Ishida:2019} show that, for SNIa classification, using only 12\% of the training set, the AL approach doubles purity with respect to the non AL approach. More recently, \cite{Leoni:2022} adapted this method to early SNIa classification using real data from ZTF alerts. This work uses photometric data and public labels from other follow-up programs, thus, it does not prioritise follow-up in real-time. Recently, the RESSPECT group has created a recommendation system for spectroscopic follow-up based on AL targeting SNIa cosmology with simulations \citep{Kennamer:2020, Malz:2023}.

Here we present for the first time an AL approach for training set optimisation in real-time with spectroscopic follow-up. This paper focuses on optimising spectroscopic training sets for early SNe Ia classification using the ZTF public stream with \fink. While this work can be extended to other transient and variables, SNe Ia is a good benchmark for the following reasons. First, SNe Ia are widely used in cosmology to constrain the expansion of the Universe \citep{DES:2024}. This means they are of great interest for accurate photometric classification. Second, SNe Ia early light-curves, within a couple of days of the explosion, are not properly modelled and understood. This is an active area of research comprising the retraining of the SALT SNIa model \citep{Guy:2007,Kenworthy:2021}, and coordinated observation programs such as Dark Energy Bedrock All Sky Survey (DEBASS; Brout et al. in prep.). These programs rely on identifying potential SNe Ia early on to obtain spectra for improving the Spectral-Energy Distribution (SED) model. 

The goal of this work is to optimise follow-up in the context of improving training sets for early SN Ia classification. This objective is very different from that of obtaining high-purity SN Ia samples with complete light-curves for cosmology as done with {\sc SuperNNova} \citep[SNN;][]{Moller:2020} for the Dark Energy Survey cosmology analysis \citep{DES:2024, Moller:2022}. First,
in this work we aim to classify early SNe Ia which is a challenging task as our current SED models, which are used for simulations, are not complete. For DES we classify complete light-curves which are much better understood and thus simulated. SNN is a non-parametric Deep Learning model that requires large training sets to achieve its full performance and thus uses simulations. Second, the accuracy and purity requirements are very different in both analyses. In particular, for cosmology, we aim for high purity, thus contamination must be small, not sensitive to training set perturbations and thoroughly modelled. SNN has been shown to meet these requirements in the DES analysis with accuracies >98\% \citep{Moller:2022,Vincenzi:2021}. In the case presented here, the relative improvement in accuracy between the initial and the final state is more important than the final overall accuracy -- which, given an informative training sample, can certainly be improved by different choices of algorithm and hyper-parameters. In summary, the approach presented here does not seek to emulate that used for SN Ia cosmology but instead aims to optimise spectroscopic follow-up to improve training sets and potentially enhance our SED models used for simulations.

This paper is organised as follows. First, we introduce the data used for this work in Section~\ref{sec:ztf}. In Section~\ref{sec:ML}, we present the early SN Ia machine learning algorithm, feature extraction method and evaluation metrics. To optimise follow-up we use an Active Learning strategy in Section~\ref{sec:AL} which includes candidate selection, communication, follow-up observations and retraining of the ML model. We present the results from the \fink\ AL loop in Section~\ref{sec:ALfink}, including an comparison of ours with different follow-up strategies. With the advent of the Rubin Observatory LSST, we propose improvements for this follow-up strategy in Section~\ref{sec:Rubin}.

\section{ZTF data}\label{sec:ztf}

We make use of the public alert stream from ZTF \citep{Bellm:2019} accessed using the \fink\ broker \citep{Moller:2021}. The ZTF alert stream is an unfiltered, 5-sigma photometry stream in the $g$ and $r$ bandpasses. Each night, \fink\ ingests and processes in real-time the alert packets sent by the telescope.

We used the photometric information contained in alerts received by \fink. These correspond to the object identification, the time of observation, the 5-sigma detection magnitude from PSF-fit photometry and the error in magnitude, for each of the ZTF filters. \fink\ also provides a cross-match with the SIMBAD database within a 1'' radius \citep{SIMBAD:2000}.

In this work we make use of three samples of ZTF alerts: an initial training sample, a testing sample and a loop sample.

The initial training and testing samples are selected from the data used in \cite{Leoni:2022} from the public ZTF stream between November 2019 to March 2020. All the light-curves have an assigned type from either the SIMBAD database or the Transient Name Server\footnote{\url{https://www.wis-tns.org/}} . As shown in Table~\ref{tab:types_Leoni}, the initial training sample consists of 30 light-curves of SNe and other variable events; while the testing sample is composed by $2,340$ SNe. We highlight that the large number of SNe Ia in the test sample is a product of the astrophysical rates together with the follow-up strategies for different surveys that are reported in TNS.

The loop sample is composed of public alerts received in real-time in \fink\ from the ZTF survey from September 2023 to 2024. We will further filter this sample, constrain it to our follow-up programme observing seasons, and use it chronologically as presented in Section~\ref{sec:preprocessing}.

\begin{table}[]
\begin{tabular}{lrr}
\toprule
Type & Initial training sample & Testing sample \\
\midrule
AGN & 1 & 0 \\
C* & 2 & 0 \\
EB* & 3 & 0 \\
SN Ia & 13 & 1587 \\
Mira & 1 & 0 \\
QSO & 7 & 0 \\
RRLyr & 4 & 0 \\
SLSN-I & 1 & 63 \\
SLSN-II & 0 & 30 \\
SN & 0 & 6 \\
SN I & 0 & 20 \\
SN II & 3 & 341 \\
SN II-pec & 0 & 2 \\
SN IIP & 1 & 51 \\
SN IIb & 0 & 24 \\
SN IIn & 1 & 106 \\
SN Ib & 1 & 21 \\
SN Ib-pec & 0 & 1 \\
SN Ibn & 0 & 7 \\
SN Ic & 0 & 73 \\
SN Ic-BL & 0 & 8 \\
Star & 1 & 0 \\
Variable star & 1 & 0 \\
\bottomrule
\end{tabular}
\caption{SIMBAD types of the samples used for initial training and testing. The samples are subsets of those in \cite{Leoni:2022}.}
    \label{tab:types_Leoni}
\end{table}

\section{Machine learning classification}\label{sec:ML}

In this work we use a Machine Learning (ML) algorithm to obtain classification probabilities for ZTF alerts. These probabilities are then assessed with an Active Learning approach to optimise follow-up observations (Section~\ref{sec:AL}). We introduce the ML algorithm in Section~\ref{sec:rf}, the method to extract features from light-curves for algorithm input in Section~\ref{sec:features} and the metrics used to evaluate its performance in Section~\ref{sec:metrics}. Both are based in the approaches presented in \cite{Ishida:2019} and adapted to the \fink\ broker by \cite{Leoni:2022}. We then train the initial version of the classifier in Section~\ref{sec:initialstate}.

\subsection{Algorithm}\label{sec:rf}

The classification algorithm used in this work is a Random Forest classifier \citep{Ho:1995}. A Random Forest is an ensemble method that uses a number of decision trees, trained from different sub-samples of the training set. The classification output is determined by the majority of the predictions for the trees in the ensemble.

Random Forests have been successfully used for a variety of astronomical classification problems \citep[][]{Ishida:2013,Moller:2016,Leoni:2022}. Furthermore, ensemble methods have been shown to be more robust in classification also for Deep Learning applications in astronomy \citep[][]{Moller:2022}.

Moreover, for the specific purpose of this project, its most important quality is the sensitivity to small changes in the training set. Decision trees divide the parameter space into small regions around each object in the training sample. Thus, in the small training data regime, it quickly adapts classification results when faced with a small number of new labels. This is a crucial feature for any classifier which needs to work within an active learning framework.

All Random Forest models trained in this work were constructed using 1000 trees, to ensure good convergence in a reasonable time. The algorithm is trained to provide high accuracy classification in the binary problem early SNe Ia vs non SNe Ia, providing as output a probability of the light-curve being an early SNe Ia, $P_{Ia}$ .

\subsection{Feature extraction} \label{sec:features}

To extract features from multi-band light-curves, we follow the procedure descried in \cite{Leoni:2022}. First, ZTF magnitudes are converted to SNANA \citep{Kessler:2009} flux units, $f$,  and its corresponding error, $\Delta f$,
\begin{eqnarray}
    f & = & 10^{-0.4m + 11}, \\
    df & = & 10 ^{10} * \alpha * dm * \exp{\left(-\frac{\alpha m}{10} \right)}
    \label{eq:fluxcal}
,\end{eqnarray}
where ${\alpha = 4 \times \ln{10} \sim} 9.21034$. Then, observations in each filter were independently fitted with a sigmoid, $S$. This sigmoid fit at a given time $t_i$ is described by: 
\begin{equation}
{S}(t_i) = \dfrac{c}{1 + e^{-a (\Delta{t}_i - b) }},
\label{eq:fit-sigmoid}
\end{equation}
where $\Delta t_i = t_i - \min(t)$ is the observation time of the $i$-th data point since first detection. For each alert, three features in each band are obtained for the best fit values of $a$, $b$ and $c$ using the least square minimisation routine from \texttt{Scipy} \citep{Virtanen:2020}. Three other features were extracted per band: the quality of the fit $\chi^2$, the mean signal-to-noise ratio $S/N$, and the number of epochs $N$ used in the fit, as in \cite{Leoni:2022}. 

In summary, for each alert we obtain six features,
\begin{equation}
    a,b,c,\chi^2,S/N,N
\end{equation}
per band. These are used as input for the classification algorithm introduced in Section~\ref{sec:rf}. We require at least one filter with 3 detections to extract features from a light-curve.

\subsection{Evaluation Metrics} \label{sec:metrics}

We evaluate the performance of the classification using the metrics introduced by the Supernova Photometric Classification Challenge \citep{Kessler:2010PC}. 

\begin{eqnarray}
{\rm accuracy} & = & \frac{C_{\rm Ia} + C_{\rm non-Ia}}{N}, \\
{\rm efficiency} & = & \frac{C_{\rm Ia}}{N_{\rm Ia}}, \\
{\rm purity} & = & \frac{C_{\rm Ia}}{C_{\rm Ia} + W_{\rm non-Ia}}, \\
{\rm fom} & = & {\rm efficiency} \times \frac{C_{\rm Ia}}{C_{\rm Ia} + W^{*} \times W_{\rm non-Ia}},
\end{eqnarray}
where \texttt{fom} stands for figure of merit, $C_{\rm Ia}$ is the number of correctly classified Ias, $C_{\rm non-Ia}$ denotes the number of correctly classified non-Ias, $N$ represents the total number of objects in the target sample, $N_{\rm Ia}$ is the total number of Ias in the target sample, $W_{\rm non-Ia}$ is the number of wrongly classified non-Ias, and $W^*=3$ \citep{Kessler:2010PC} is a weight which penalizes false positives. 

These metrics were used only to evaluate the performance of the trained ML algorithm in the independent testing sample and they were not used in the decision-making process of the Active Learning Loop.

\subsection{Initial model}\label{sec:initialstate}

We train an initial model using the initial train sample of 30 light-curves from the ZTF survey introduced in Section~\ref{sec:ztf}. These light-curves are a subset of the sample by \cite{Leoni:2022} and contain both SNe and variables. The algorithm is trained to select normal early SNe Ia vs non SNe Ia.

The initial trained RF model has an accuracy of 0.47, an efficiency of 0.26, purity 0.87, and a fom 0.18. It provides a classification which has high purity but low efficiency for SNe Ia. This will be our starting point for the Active Learning strategy in this work.

\section{Active Learning}

In cases where unlabelled data is abundant but labelling is expensive and/or time consuming, Active Learning \citep{Settles:2012} provides a way to optimise the acquisition of labels that will improve the model when incorporated to the training set. 

\begin{figure*}
        \centering
        \includegraphics[width=\linewidth]{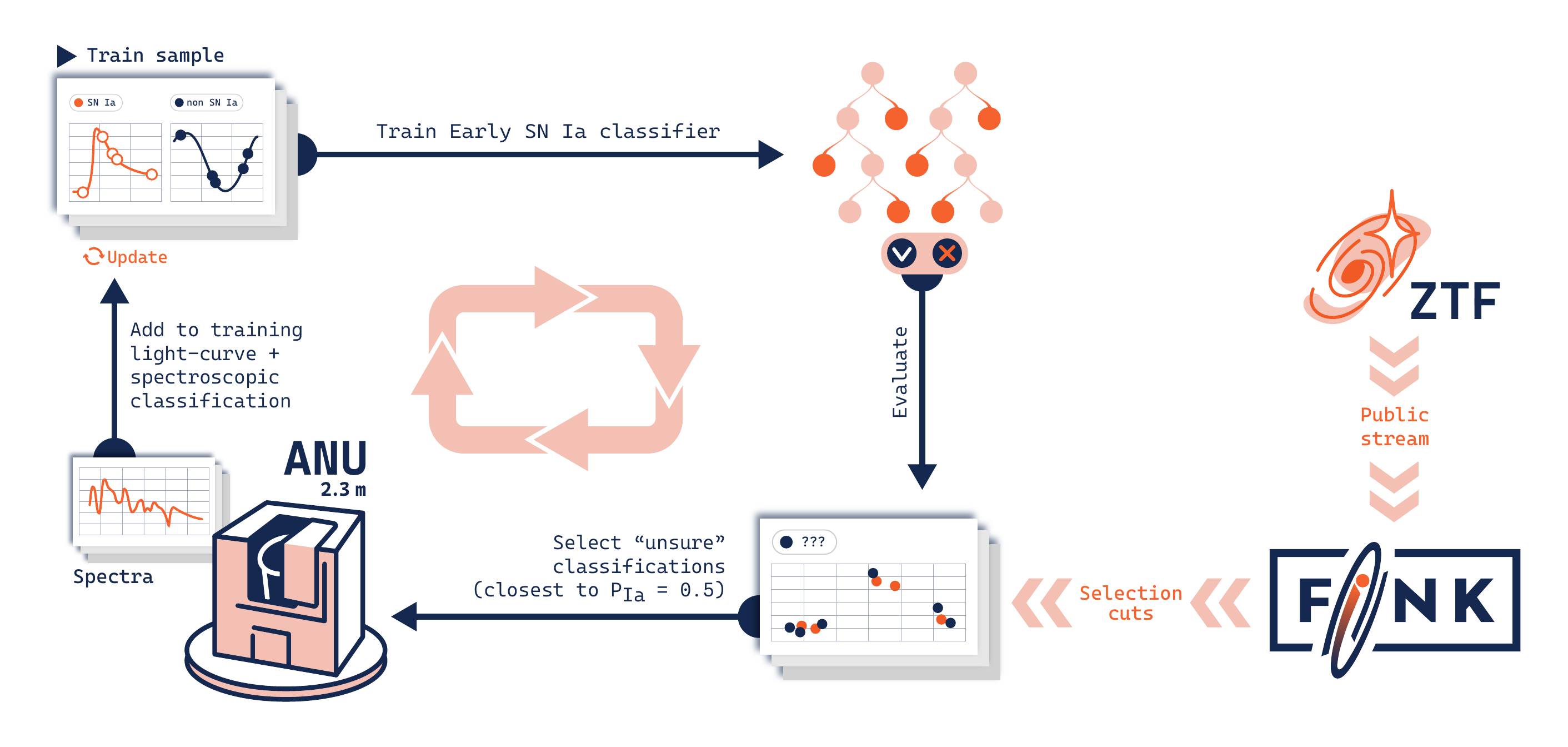}
        \caption{Active Learning loop schema. The loop starts with the Initial train sample which is used to train the Early SN Ia classifier, this algorithm is then applied to alerts processed by the Fink broker from the ZTF public stream. We select alerts which obtain the closest $P_{Ia}$ to 0.5 and schedule spectroscopic follow-up with the ANU 2.3m if they have no spectroscopic classification. Once a label is obtained, we add the light-curves and labels for the selected events to the training set. The loop is repeated during the observing period.}
        \label{fig:alloop}
    \end{figure*}

For our work, we perform an observation campaign between September 2023 and August 2024. In real time, as indicated in Figure~\ref{fig:alloop}we select follow-up candidates from ZTF data (Section~\ref{sec:preprocessing}) with an Active Learning strategy (Section~\ref{sec:AL}). These candidates are selected nightly and communicated through a \fink\ bot (Section~\ref{sec:finkbot}). In the subsequent night, we schedule follow-up using the ANU 2.3m telescope (Section~\ref{sec:fup}). Once a spectroscopic classification is obtained, new labels and light-curves are incorporated to the training set (Section~\ref{sec:loop}) and their performance assessed. This is a chronologically ordered loop.

\subsection{ZTF alerts selection}\label{sec:preprocessing}

To select early SNIa-like events, we use a modified filtering inspired by the early SNIa substream in Fink \citep{Moller:2021,Leoni:2022}. We use the loop sample defined in Section~\ref{sec:ztf}, which contains alerts within our follow-up programme dates (see Section~\ref{sec:fup} and Table~\ref{tab:observingseasons}). Asides from the date constraints, we select alerts which:
\begin{itemize}
    \item Are not known variable stars or AGNs identified by cross-matching their coordinates with SIMBAD database and a 1'' radius.
    \item Have less than 20 days between their first and latest detection.
    \item Have less than 20 measurements.
\end{itemize}

We construct a light-curve using all detections available in the public stream and then obtain light-curve features for each filter individually using the fit, eq.~\ref{eq:fit-sigmoid}, in Section~\ref{sec:features}. 
We then obtain a classification probability for each of these candidates.

\subsection{Active Learning Strategy}\label{sec:AL}

In the context of classification, instead of selecting high probability candidates for a high efficiency or pure sample, an Active Learning approach will select the most uncertain classifications for label acquisition, in our case spectroscopic follow-up. 

In \cite{Ishida:2019, Leoni:2022} they used Uncertainty Sampling as the Active Learning strategy. Uncertainty sampling \citep{sharma2017} identifies the objects closest to the decision boundary between classes at each iteration. As our classification is a binary one, the selection would correspond to the object with a classification probability closest to $P_{Ia} \sim 0.5$. See, e.g., \cite{MacKay:2003}. For multi-class classifiers potential selection strategies can be found in \cite{Xue:2019}.

For this work, in the case of our binary classification, we find that candidates with probability between 0.4<$P_{Ia}$<0.6 are quite rare. Given further constraints to the follow-up observations such as magnitude limit and weather condition, at each iteration, we choose to select the closest 10 alerts to $P_{Ia}=0.5$.

\subsection{Fink bot}\label{sec:finkbot}

In the following, we summarise three methods currently available in the \fink\ broker to access filtered candidates. We then detail the implementation of the \fink\ AL loop bot.

In the \fink\ broker, filtered candidates can be automatically communicated in the following ways:
\begin{itemize}
    \item Substream through the \fink\ client: receiving full-content filtered candidates as they are processed.
    \item Substream through a \fink\ bot: receiving personalised notifications based on filtered candidates using instant messaging applications such as Slack or Telegram.
\end{itemize}

The delay between an observation is received, ingested, processed and filtered by \fink\ has currently a median delay of $78s$. The 10 and 90 percentiles of this delay are $39s$ and $123s$. This amounts to a delay between less than a minute and up to two minutes for accessing promising candidates after their data has been received.

The use of instant messaging applications, such as Slack or Telegram, has proven to be very convenient for users in Fink, and the implementation of bots has been made easy thanks to the availability of public and interoperable API. Users can then receive information within the applications they already use regularly, without the need to install a new tool or navigate to a separate website.

We note that there is a third option to access candidates, by manually querying the \fink\ Science Portal or using its API\footnote{\url{https://fink-broker.readthedocs.io/en/latest/services/search/getting_started/}}. However, the database used in this service is only populated at the end of the night and thus has usually several hours delay to access the promising candidates.

For the \fink\ AL loop, we choose the substream processing with a Slack bot. We introduce an additional delay, as we choose to communicate the 10 candidates closest to $P_{Ia}=0.5$ for the night. The whole operation takes typically less than a minute for an entire night (about 200,000 alerts), dominated by the classification task.

These 10 candidates are automatically communicated via an instant messaging application by a bot. The Slack message, which is customizable, includes the most recent thumbnail and full light-curve, together with a link to the object in the portal. Candidates are then visually inspected by a community of users and some are scheduled for follow-up observations if they are observable by the ANU 2.3m (e.g. airmass, telescope availability and magnitude limit criteria).

\subsection{Follow-up observations} \label{sec:fup}

We perform follow-up observations with the ANU 2.3m telescope located in Siding Spring Observatory in Australia. We had four distinct follow-up periods due to weather and technical constraints as shown in Table~\ref{tab:observingseasons}.

We use the Wide Field Spectrograph \citep[WiFeS, ][]{Dopita:2007} which has a field of view of 25 × 38 arcseconds. We obtain $R=3000$ resolution spectra using the R3000 and B3000 gratings, for the red and blue arms of the spectrograph and the RT560 dichroic. The spectra obtained cover a wavelength range of 3500 - 9000Å.

\begin{table}[]
    \centering
    \begin{tabular}{c|c|c}
       Follow-up season & Start  &  End \\
       1 & 25/09/2023 & 23/10/2023 \\
       2 & 25/02/2024 & 29/02/2024 \\
       3 & 04/04/2024 & 02/06/2024\\
       4 & 24/06/2024 & 15/08/2024 \\
    \end{tabular}
    \caption{Date ranges for alerts used for this work.}
    \label{tab:observingseasons}
\end{table}

We observed the candidates using the Nod \& Shuffle (N\&S) mode to obtain a better signal-to-noise ratio. Depending on the magnitude of the transient and the moon phase, we observed either two or three N\&S $1200s$ exposures ($600s$ on target, $600s$ on sky). If possible, we included the host-galaxy in the field-of-view to obtain its spectra simultaneously.

If a host-galaxy spectrum is obtained, we use the software MARZ \citep{Hinton:2016} to obtain its redshift. If the host is not within the field or magnitude limits of our spectra, we search available catalogues for a redshift estimation whether spectroscopic or photometric.

We then classify the candidates using their spectra. For this, we use SUPERFIT classification package written in IDL \citep{Howell:2005}. SUPERFIT compares the given spectra to a set of transient and host templates. We provide an input spectrum and a redshift range or host redshift if available. The input spectrum is sequentially compared to each of these templates while iterating through the redshifts if needed, reddening corrections and different levels of host-galaxy contamination. We then visually inspect the top matches to obtain a classification. Spectra with a quality classification are then reported to the IAU Transient Name Server.

\subsection{AL loop}\label{sec:loop}

Once a spectroscopic classification is obtained, the labelled light-curve is added to the training and the machine learning model is retrained. This new retrained model is deployed and it is used to obtain the next night's candidates for follow-up. The iterations continue with the goal of improving  the classification algorithm.

The labelled light-curve used for feature extraction and thus retraining contains only photometry until the follow-up request. For the following results, we choose to add the obtained label the day following when the follow-up was requested. This is to homogenise the system, as in reality, as discussed in Section~\ref{sec:automatic_fup} the label can be acquired days after it was selected by the bot. This simplification does not alter the results found.

\section{Active Learning follow-up with \fink} \label{sec:ALfink}

We present here the results of the application of the Active Learning strategy for follow-up with \fink\ using the ZTF public alerts. First, we show the performance evolution of the \fink\ AL loop chronologically in Section~\ref{sec:alperformance}. Then, we analyse how the AL approach optimises follow-up resources in Section~\ref{sec:optimisingfup}. We conclude by exploring the differences between the \fink\ AL follow-up targets and other follow-up strategies in Section~\ref{sec:fup_targets}.

In the following we will compare our \fink\ AL loop strategy with labels available in TNS: i) those reported by the ZTF group (ZTF); ii) all TNS available labels containing different follow-up strategies (TNS). The latter, TNS, contains ZTF reported classification together with the reports of all other groups.

\subsection{ML algorithm performance through time} \label{sec:alperformance}

Starting from the initial trained model in Section~\ref{sec:initialstate}, we apply the filtering described in Sections~\ref{sec:preprocessing} and \ref{sec:AL} to select follow-up candidates from the ZTF public alert stream with \fink. We scheduled follow-up with the ANU 2.3m and obtained labels if the spectra had a signal-to-noise ratio high enough for classification (Section~\ref{sec:fup}). Chronologically, new labels and their respective light-curves (with photometry up to the follow-up day) were incorporated to the training set, the ML model was re-trained, and its performance evaluated on the independent testing set. 

\begin{figure*}
    \centering
    \includegraphics[width=\linewidth]{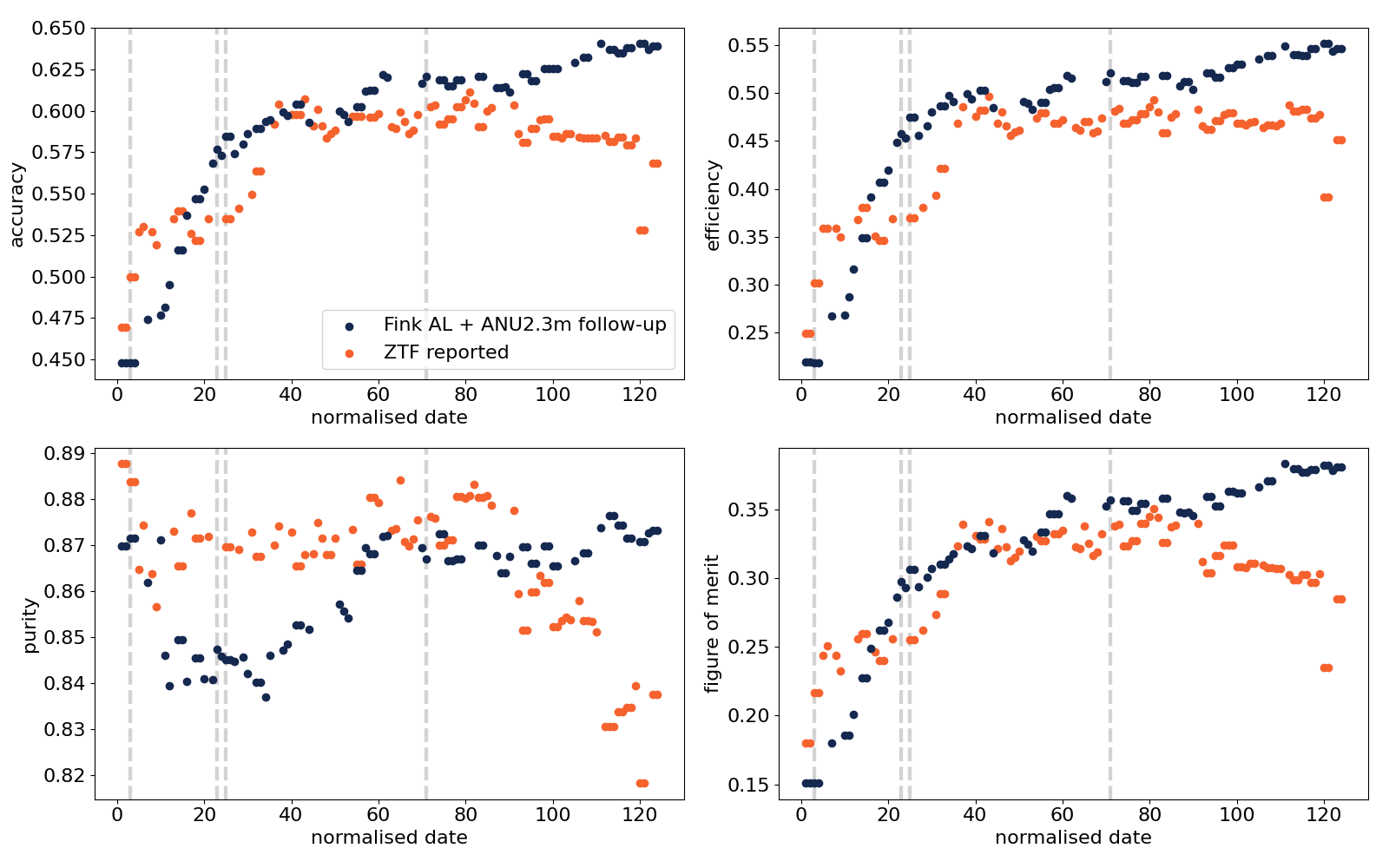}
    \caption{Evolution of classification metrics as a function of the number of normalised follow-up date from the \fink\ AL strategy and using all TNS reported ZTF classifications. Metrics are evaluated on the independent testing set every date that a new label is available. The grey vertical lines show the observing seasons boundaries.}
    \label{fig:metrics_fink_ztf_date}
\end{figure*}

We see the evolution of the ML algorithm in Figure~\ref{fig:metrics_fink_ztf_date} as a function of a normalised date. The normalised date skips epochs between observing seasons to show a continuous  evolution. The metrics are evaluated in the same independent testing sample each time the algorithm is retrained with the acquired labels from the \fink\ AL follow-up strategy. We compare, for the same date range, a loop with all labels reported by ZTF in TNS. For the latter, to select the portion of the light-curve to use in the loop consistent with \fink\ AL early follow-up selection, we check the difference between \fink's follow-up date and their first detection. We find that most of our follow-up targets are chosen 9 days after their discovery or first detection date. We use this time span to select the TNS reported light-curves. Thus, the TNS light-curves contain only photometry for 9 days after discovery at all points in loop.

Both strategies start at the same initial state, which was defined by the initial 30 objects (15 SNIa, 15 others, picked at random from the sample curated by \citet{Leoni:2022}). In this state, efficiency is low and purity is high. This is an indication of the low diversity among the Ias present in this initial set. Since a low number of cells in the forest are identified as SNIa, a very low number of Ia candidates are identified, hence the low efficiency and high purity which results in a low figure of merit. As more informative samples are added to the training, this imbalance is corrected. Figure \ref{fig:metrics_fink_ztf_date} shows that
while the \fink\ and ZTF loops have similar performances with the first acquired labels, after $\approx 50$ loops the \fink\ strategy consistently increases its performance while the ZTF one does not, illustrating the fact that later objects in the ZTF sample do not contain information which was not already present in training. We highlight that in the last iterations, the ZTF loop performance strongly decreases due to two light-curves which do not have a successful feature extraction. 

For the first 15 loops, we see an increase in efficiency in the SN Ia classification and a decrease in the purity of the sample. This is due to an increase in contaminant selection, from other SNe core-collapse and super-luminous types. The main contaminants found are type II and Ic SNe followed by SLSN-I. After the 20th loop, the \fink\ AL model strongly  reduces the SLSN and in a lesser extent core-collapse SN contamination.

As our classifier improved we deployed it in the \fink\ broker to select both new follow-up candidates and also, using a high probability cut, select promising early SNe Ia. Some of these SNe Ia have been further analysed and incorporated to analyses such as the Dark Energy Bedrock All Sky Survey (DEBASS; Brout et al. in prep.).

\subsection{Optimising follow-up resources}\label{sec:optimisingfup}

Spectroscopic follow-up resources are scarce for current and future surveys. In this Section we explore whether the \fink\ AL approach is efficient, expressed as the number of spectra, for label acquisition to improve the performance of the classification algorithm.

\begin{figure*}
    \centering
    \includegraphics[width=\columnwidth]{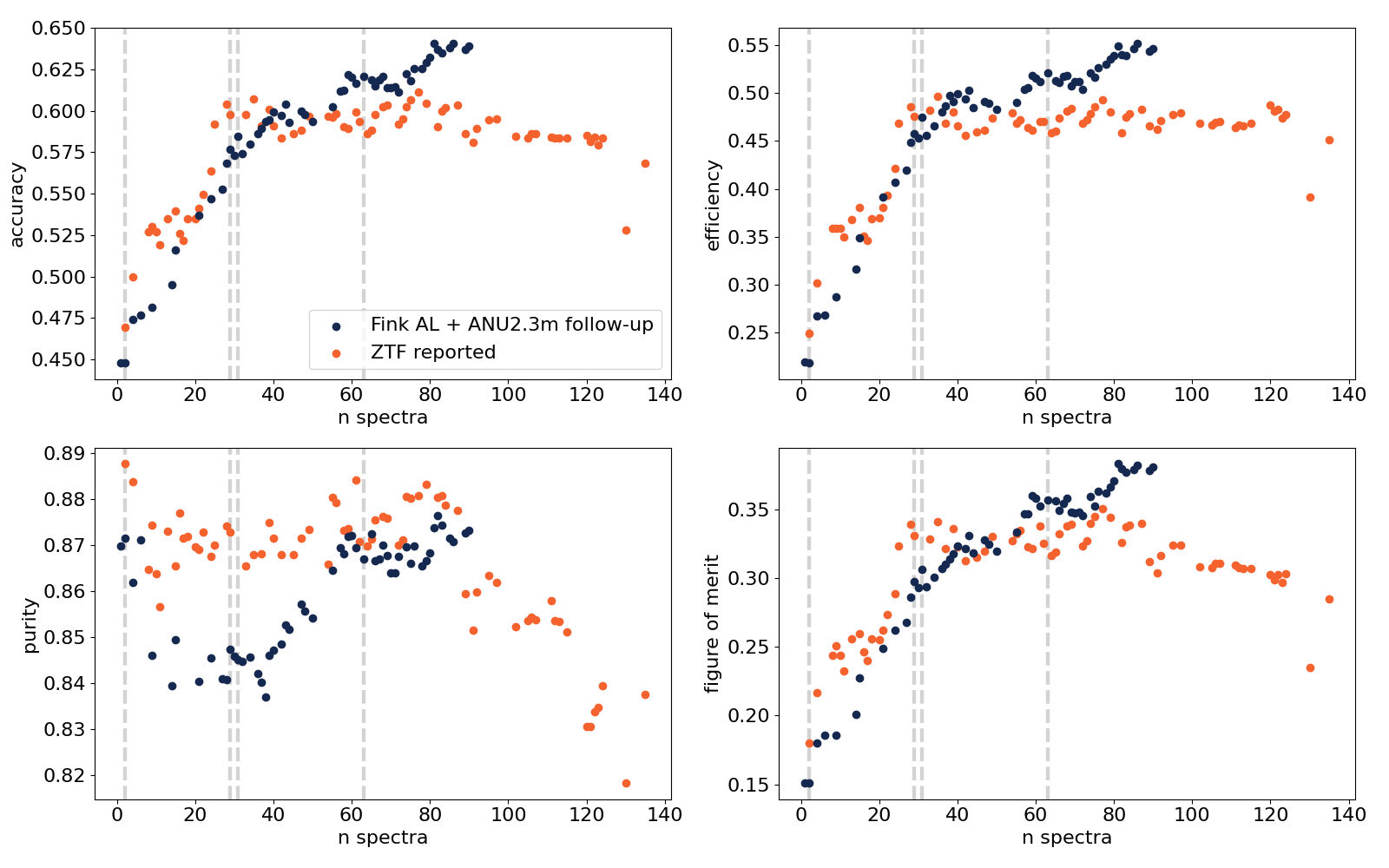}
    \caption{Evolution of classification metrics as a function of the number of spectra (n spectra) taken for the {\sc Fink} AL strategy and using all TNS reported ZTF classifications. Metrics are evaluated on the independent testing set every date that a new label is available. We used a normalised date which ignores breaks between follow-up observing seasons. The grey vertical lines show the observing seasons boundaries.}
    \label{fig:metrics_fink_ztf_nspectra}
\end{figure*}

In Figure~\ref{fig:metrics_fink_ztf_nspectra} we compare the evolution of metrics with \fink\ AL strategy with ZTF follow-up as a function of the number of spectra acquired. This differs from a date evolution, as several spectra can be acquired in a single observing night. For the same date range, the \fink\ AL strategy has acquired around half of the spectra of ZTF and obtained better performance. This shows that an AL strategy can optimise follow-up resources for improving photometric classification training sets.

We further explore the optimisation of follow-up resources by comparing the performance of our \fink\ AL strategy with respect to the full TNS reported classifications in this period. To achieve a similar performance, we require $\ge 127$ TNS classified spectra vs. 92 \fink\ ones (equivalent to a reduction of $25\%$). In telescope time, taking our observing configuration of $1,200~s$ per target, this represents around one and a half nights of observation out of 5.3 nights that are saved by this approach if all targets are observed. As we will discuss in the next Section~\ref{sec:fup_targets}, we are not 100\% efficient in acquiring spectra, e.g. some spectra has not enough signal to obtain a reliable classification. Thus, the number of scheduled time is larger (around a factor 2) than the number of spectra in the loop multiplied by the observing time.

\subsection{Follow-up targets }\label{sec:fup_targets}

With the \fink\ AL strategy we identify 178 follow-up candidates. 110 of these candidates have a label from our ANU 2.3m follow-up, catalogues or reported in TNS (68, 7 and 35 respectively).

We find that most of the follow-up candidates are selected 9 days after first detection is reported. This is influenced by the need of at least 3 detections in a given filter to extract features and the observing cadence of ZTF. We explore a potential improvement to target earlier light-curves in Section~\ref{sec:Rubin}.

\begin{figure}
    \centering
    \includegraphics[width=\linewidth]{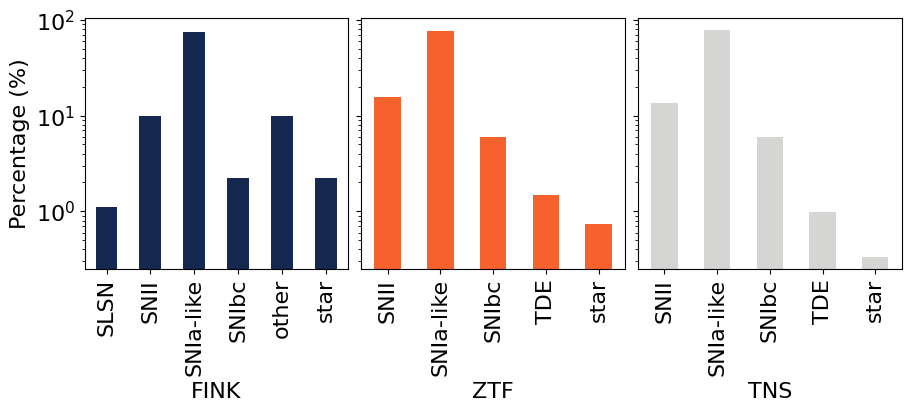}
    \caption{Spectroscopic classes for follow-up candidates in the \fink\ AL loop. We show from left to right panels \fink, ZTF and all TNS spectroscopic classifications. The percentage of SN families is similar to all strategies except for SLSN and other non-SN types of transients characterised by \fink.}
    \label{fig:ALloop_types}
\end{figure}

After quality cuts and post-processing, 90 enter into the loop. Their final classification is shown in Table~\ref{tab:fink_classes}. We compare our follow-up candidates with those obtained from TNS in Figure~\ref{fig:ALloop_types}, grouping peculiar SNe with their normal classes. While the percentage of SN types between the different follow-up targets is similar, with \fink\ AL we spectroscopically follow-up a SLSN and other transients which include microlensing events and CV.

\subsubsection{Candidate brightness}

We find that the \fink\ AL follow-up candidates have a median magnitude of $19.5 \pm 0.7 $ when appearing in the bot, similar to follow-up candidates from ZTF and all TNS. We highlight that the depth achieved by the Nod \& Shuffle mode with WiFeS allows us to type early SNe Ia, such as ZTF24aaiypmp\footnote{\url{https://fink-portal.org/ZTF24aaiypmp}} for which we obtained spectra at $19.5 ~ mag$. ZTF24aaiypmp was classified as a SNe Ia between -11 and -13 days before maximum light. 

\begin{figure}
    \centering
    \includegraphics[width=\linewidth]{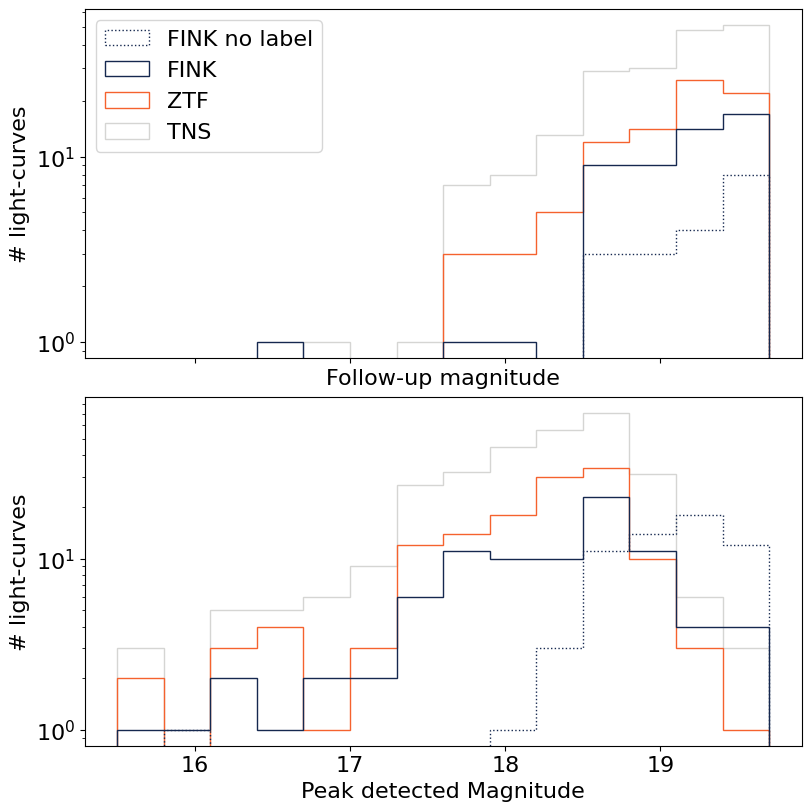}
    \caption{Magnitude distribution of candidates when triggering follow-up (top panel) and detected peak (lower panel). We show the distribution for \fink follow-up program with successful spectroscopic classification in solid lines and without in dotted lines. We also show the magnitude distribution for ZTF classified and all TNS reported spectroscopic samples during our \fink program.}
    \label{fig:peak_fup_mag}
\end{figure}

From the 178 candidates, 68 have a label from ANU 2.3m, 35 from TNS before we acquired spectra, 7 are catalogued stars, and 68 candidates were not classified. Most of these missing classifications are due to weather delays or prioritisation in the telescope (thus the SN faded away before obtaining its spectra). A small portion had spectra acquired which was not enough signal-to-noise ratio for classification. These are found to be mostly faint sources with a median magnitude $ 19.5 \pm 0.5$ when appearing in the bot. A posteriori, we check the peak magnitude of these events, as shown in Figure~\ref{fig:peak_fup_mag}. We find that most of the unlabelled candidates have a peak detected magnitude around $19.2 \pm 0.6$. Thus, they are faint SNe or other transients which peak shortly after the follow-up was triggered.

\subsubsection{Host-galaxies}
In this analysis, we did not use any host-galaxy information in the ML algorithm. In this Section, we inspect a posteriori the host-galaxy properties of followed-up candidates.

To identify the candidates host-galaxies, we use the Legacy Surveys DR10 \citep{Dey:2019} accessed through the NOIRLab Archive. We query a 1' radius around the SN to select potential hosts. For each candidate host, we compute the directional light radius (DLR) as in \cite{Sullivan:2006}. This method computes a dimensionless distance (dDLR) for each potential host-galaxy measured between the SN position and the centroid of the galaxy, in angular distance, normalised by the galaxy size in the direction of the SN, also in angular distance.

Potential hosts only include those that hat have a Legacy value for the radius, additionally we require that they have dDLR<4 as in \cite{Qu:2024}. We then apply additional cuts using available information from LS DR10 to eliminate multiple host candidates from spurious detections as well to ensure the good galaxy profile has been used. 
\begin{itemize}
\item To ensure only galaxies are selected, we apply $0<$excess\_factor$<2.5$. Excess\_factor is the ratio of fluxes ($I$) from Gaia bands ($BP$, $RP$ and $G$) computed as $(I_{BP} + I_{RP})/I_{G}$.
\item To minimise contaminated photometry from blended sources we require $fracflux>1$ for any LS band. The $fracflux$ measures the the profile-weighted fraction of flux from other sources divided by the total flux.
\end{itemize}

The host is selected as the closest host-galaxy, equivalent to the one with minimum dDLR value, that passes the cuts. We find that on average our \fink\ followed-up SNe are in slightly fainter host-galaxies than those followed up by ZTF or reported in TNS. The magnitude range varies on the colour in a range of $0.4-1 ~ mag$ fainter but within an standard deviation of the distributions.

\begin{table}[]
    \centering
    \begin{tabular}{lr}
type & number \\
\midrule
II & 8 \\
IIb & 1 \\
Ia & 62 \\
Ia-91T-like & 2 \\
Ia-pec & 2 \\
Iax[02cx-like] & 1 \\
Ibn & 1 \\
Ic-BL & 1 \\
SLSN & 1 \\
featureless & 2 \\
microlensing & 2 \\
other & 5 \\
star & 2 \\
\bottomrule
\end{tabular}
    \caption{\fink\ targets in the AL loop with types acquired with the ANU 2.3m spectra. Featureless and other indicate spectra which have no features consistent with a SN.}
    \label{tab:fink_classes}
\end{table}

\section{Improvements for Rubin LSST}\label{sec:Rubin}

The Vera C. Rubin Observatory is expected to obtain up to 10 million alerts every night during its 10-year Legacy Survey of Space and Time \citep[LSST;][]{ldm612}. Within these millions detections, there will be rare transients and early SNe Ia. \fink\ will be processing the alert stream of Rubin LSST in real-time for the next decade.

Spectroscopic follow-up will be led by the 4MOST Time-Domain Extragalactic Survey \citep[TiDES; Frohmaier et al. in prep.,][]{Swann:2019} with an emphasis on SNe Ia, AGNs, transients and their host-galaxies. However, other facilities will be following-up spectroscopically other interesting transients and variables for a wide science community. Even with additional surveys, we will be unable to get spectroscopy of more than a handful percent of the discovered transients.

As photometric classification will be key to fully exploit the transient sample of Rubin, it is paramount that we optimise follow-up resources to characterise those classes that can substantially improve our photometric classification training. This will be particularly important for rare and new transients for which our training samples are a small number of spectroscopically classified light-curves. Importantly, given the sheer volume of data from LSST, it will be crucial to optimise resources for  spectroscopic follow-up.

In the following we discuss two potential improvements for Rubin deployment targeting: earlier light-curves and a more automatic follow-up.

\subsection{Targeting earlier light-curves}

One of the main limitations of our classification method, is that we require at least 3 detections in a given filter to extract features for the ML classification algorithm. Given the ZTF observing strategy, this translates into $\approx 9$ days post first detection for an event to be a follow-up candidate. For Rubin LSST this time window will be much larger depending on which field the observations are done (e.g. Deep drilling fields vs Wide Fast Deep survey).

For observing strategies were the delta time between photometric measurements is $\le 3$ days but in different filters, an option would be two use inference methods such as  {\sc Rainbow} \citep{Russeil:2024}. This method assumes a black-body and a temperature evolution function to enable simultaneous multi-band light curve fitting. Another option would be to treat all filters together, with the drawback that colour information would be lost which can be crucial for classification.

\begin{figure}
    \centering
    \includegraphics[width=\linewidth]{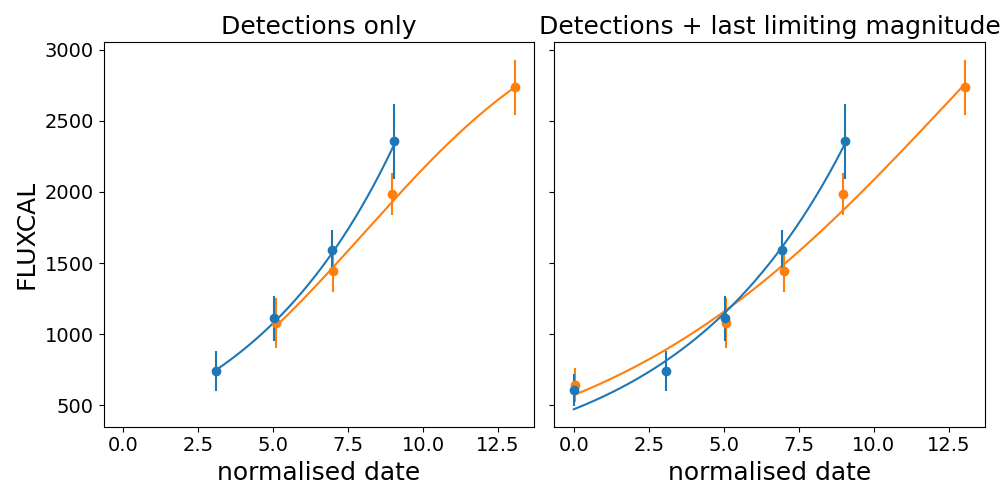}
    \caption{Example of early light-curve of candidate ZTF23abdhvou and its feature extraction. In the left panel we show the light-curve containing only detections. In the right panel we show the light-curve containing the last limiting magnitude for each filter and detections.}
    \label{fig:lc_wlimits}
\end{figure}

In this Section, we explore in detail another alternative. This is to incorporate the last non-detection limiting magnitude as a first measurement for the feature extraction process. If forced photometry were available in the survey, such as in the case of LSST, another option would be to add the last values found using forced photometry that is not a detection. In Figure~\ref{fig:lc_wlimits} we show an example of light-curve with and without using the limiting magnitude of the last non-detection.

Using the same follow-up candidates, we find that the loop obtains similar performances in all metrics with the addition of this extra photometric measurements
for feature extraction. 

To explore whether this approach provides earlier detection, we reduce by a day the photometry used for feature extraction, thus reducing a photometric epoch. We obtain the same number of light-curves that pass feature extraction but now with less detection information. Since we can't rerun the AL loop in real-time again using this new configuration, select new follow-up candidates and obtain spectra, we choose to use the already acquired labels and just evaluate the method's performance. We find that the metrics evolve in a similar trend, with less than 5\% variation.


For Rubin, we plan to use both strategies. First, by adding information regarding the last non-detection whether through limiting magnitude or forced photometry. Second, by incorporating colour information from the \textsc{Rainbow} framework \cite{Russeil:2024}.

\subsection{Automatic follow-up deployment}\label{sec:automatic_fup}

Spectroscopic follow-up can only be achieved while the candidate is visible and above the limiting magnitude of the instrument. Transients are generally short lived and to efficiently trigger follow-up it is necessary rapid follow-up with facilities that can observe them shortly after they occur. Recent work has shown that prioritisation of follow-up candidates can be done using \fink\ alerts with data from other telescopes and brokers to select follow-up candidates \citep{Sedgewick:2025}. Additionally, for AL applications, it will be necessary to automatically reduce the data and incorporate labels as quickly as possible to ensure the optimisation of further follow-up. 

A first improvement would be to select follow-up candidates based on the real-time alerts. These would entail a delay of $\approx 1 ~minute$ between the reception, processing and communication of the candidates within \fink\ broker as shown in Section~\ref{sec:finkbot}. Depending on the science target, it would be valuable to reduce the spectroscopic follow-up delay (e.g. fast evolving transients such as kilonovae, GRB afterglows, FBOTs).

A second improvement would be the automation of follow-up through a check on observability and robotic deployment of the follow-up facility such as the one previewed for 4MOST TiDES (Frohmaier et al. in prep). In this work, we have used the ANU 2.3m telescope located at Siding Spring Observatory (SSO). SSO is in the Southern Hemisphere and west of the major transient discovery facilities in South America such as Rubin.

The ANU 2.3m is currently a fully robotic telescope which schedules and observes targets automatically. We are currently building a Network that connects filtered streams from \fink, deploys automatically follow-up observations with the ANU 2.3m and other telescopes at SSO, and reduces data automatically. The infrastructure connecting \fink\ and the telescopes at SSO, is based on the TOM toolkit \citep{Street:2024}. 

Such a system will allow us to reduce the delay between identifying a follow-up candidate and obtaining its label. While for the purpose of this analysis we have simplified our loop by assuming that labels were acquired chronologically the next day after appearance in the bot, the reality was that it would take from $\approx 1$ to $5$ days for the observation to be done, reduction of data, and the label extracted. This delay does not change the results but would change the chronological evolution.

\section{Conclusions} \label{sec:conclusions}

In this work, we present the first real-time Active Learning application on survey data to optimise follow-up with the goal of improving photometric classification training sets. We apply this method to improve the classification of early type Ia supernovae. We make use of the ZTF public alert stream processed by the \fink\ broker to identify follow-up candidates. We perform spectroscopic follow-up observation with the ANU 2.3m WiFeS instrument located at Siding Spring Observatory in Australia.

Using the \fink\ broker infrastructure we deploy the ML classifier and filtering criteria to select the most promising follow-up candidates to improve ML training sets. The processing is done in real-time as the ZTF public alert stream is ingested and analysed. We communicate these candidates automatically through a \fink\ bot and schedule subsequent spectroscopic follow-up. We identify 177 follow-up candidates in 4 observing periods between September 2023 and August 2024. From these candidates we obtain 109 classifications, 92 are chronologically added to the training set after quality cuts.

We find that the AL strategy identifies follow-up candidates that improve the ML algorithm in a more effective way than reported classifications from the ZTF survey. This is seen chronologically as well as when comparing the number of spectra necessary to achieve a given performance. Our method, reduces the need to schedule more than one and a half nights of follow-up when using $1,200~s$ exposures, equivalent to using 90  instead of 127 spectra. 

The active learning strategy identifies candidates that are in average $\approx 19 ~ mag$, similar to the magnitude post 9 days of discovery of ZTF transients and variables. We find two main differences between our spectroscopically classified events and those reported in TNS by ZTF or other groups. First, we identify follow-up candidates that are overall fainter throughout their whole evolution. Second, we identify a SLSN and several non-SN candidates for follow-up that improve our performance, such as microlensing events and CVs. These are not typically in the simulated training sets used for ML algorithm training nor in the traditional follow-up surveys.

With the advent of Rubin LSST, it will be crucial to develop fast photometric classification algorithms to identify early, known and new classes of events. For this, we need to optimise the way we construct training sets. In this work, we have shown that AL is a good strategy for spectroscopic follow-up to improve training sets using the early SN Ia problem. The AL strategy could be particularly useful to construct training sets to improve the identification of rare transients. For example, we could improve the classification of rare transients by using a handful of observed light-curves together with potential contaminants to train a classification algorithm and apply an AL strategy. This strategy would then select both contaminants and targets for follow-up and improve the photometric classifier. This improved classifier can then be  used in and independent dataset to obtain larger samples of the target transient.

This works serves as pilot for other AL applications to improve photometric classification of, not only early SNe Ia light-curves, but new and rare astrophysical transients. With spectroscopic resources being scarce currently and in the future, the AL method is promising to optimise follow-up resources for teams aiming to fully exploit the power of future surveys such as Rubin.

\begin{acknowledgement}

We would like to thank Chris Lidman for advise on data reduction and spectroscopic classification. We would like to acknowledge Anastasiia Voloshina for the design of Figure 1 of this work.

This work was developed within the \fink\ community and made use of the \fink\ 
community broker resources. \fink\ is supported by LSST-France and CNRS/IN2P3. This research has made use of the SIMBAD database,
operated at CDS, Strasbourg, France.

Based in part on data acquired at the ANU 2.3-metre telescope. The automation of the telescope was made possible through an initial grant provided by the Centre of Gravitational Astrophysics and the Research School of Astronomy and Astrophysics at the Australian National University and through a grant provided by the Australian Research Council through LE230100063. The Lens proposal system is maintained by the AAO Research Data \& Software team as part of the Data Central Science Platform. We acknowledge the traditional custodians of the land on which the telescope stands, the Gamilaraay people, and pay our respects to elders past and present.

\end{acknowledgement}

\paragraph{Funding Statement}


This research was supported by grants from the Australian Research Council. AM is supported by the ARC Discovery Early Research Award (DE230100055). Parts of this research were conducted by the Australian Research Council Centre of Excellence for Gravitational Wave Discovery (OzGrav), through project numbers CE170100004 and CE230100016.

\paragraph{Competing Interests}

None

\paragraph{Data Availability Statement}
Spectroscopic classification tables and 
analysis code available in \url{https://github.com/Fink-analyses/Active_Learning_earlySNIa}. This repository includes the Random Forest classifier and initial training set used in this work.

\printendnotes

\printbibliography



\end{document}